\documentclass[
 aps, pra,
 amsmath,amssymb,
 11pt,
 final,
tightenlines,
 twoside,
 twocolumn,
nofootinbib,
 superscriptaddress,
showkeys,
showkeywords,
 ]
{revtex4-2}

\usepackage[T1]{fontenc}
\usepackage[utf8x]{inputenc}
\usepackage[english]{babel}
\usepackage{graphicx}% Include figure files
\usepackage{dcolumn}% Align table columns on decimal point
\usepackage{bm}% bold math
\usepackage{longtable}

\input{maik.rty}

\setcitestyle{authoryear,round}
\def\saoname{Special Astrophysical Observatory,  Russian Academy of Sciences,
              Nizhnii Arkhyz, 369167 Russia}

\def\squareforqed{\hbox{\rlap{$\sqcap$}$\sqcup$}}

\def\sq{\ifmmode\squareforqed\else{\unskip\nobreak\hfil
\penalty50\hskip1em\null\nobreak\hfil\squareforqed
\parfillskip=0pt\finalhyphendemerits=0\endgraf}\fi}

\def\utw{\smash{\rlap{\lower5pt\hbox{$\sim$}}}}

\def\udtw{\smash{\rlap{\lower6pt\hbox{$\approx$}}}}

\def\diameter{{\ifmmode\mathchoice
{\ooalign{\hfil\hbox{$\displaystyle/$}\hfil\crcr
{\hbox{$\displaystyle\mathchar"20D$}}}}
{\ooalign{\hfil\hbox{$\textstyle/$}\hfil\crcr
{\hbox{$\textstyle\mathchar"20D$}}}}
{\ooalign{\hfil\hbox{$\scriptstyle/$}\hfil\crcr
{\hbox{$\scriptstyle\mathchar"20D$}}}}
{\ooalign{\hfil\hbox{$\scriptscriptstyle/$}\hfil\crcr
{\hbox{$\scriptscriptstyle\mathchar"20D$}}}}
\else{\ooalign{\hfil/\hfil\crcr\mathhexbox20D}}%
\fi}}

\begin{document}

\selectlanguage{english}


\title{New Radio Data on Sources of the Big Trio Program for Searching for Distant Radio Galaxies}
\author{\firstname{O.}~\surname{Zhelenkova}}
 \affiliation{\saoname}
  \email{zhe@sao.ru}

\author{\firstname{A.}~\surname{Temirova}}
\affiliation{Saint Petersburg branch of SAO RAS, Pulkovskoe sh., 65, St. Petersburg, 196140~Russia}

\author{\firstname{Yu.}~\surname{Parijskij}}

\author{\firstname{N.}~\surname{Soboleva}}

\begin{abstract}

Radio sources with steep and ultra-steep spectra from the Cold Experiment surveys conducted with the RATAN-600 radio telescope formed the basis of the Big Trio program for searching for distant radio galaxies. 
With the advent of new radio, optical, and infrared sky surveys, it has become possible to conduct an in-depth study of 113 sources to determine their evolutionary status and the characteristics of their environments, as well as to analyze long-term changes in their spectral characteristics.
Based on the identified morphological and spectral features, a fifth of the sources show signs of one of the evolutionary phases -- initial, fading, or resumption of activity. 
Based on morphological features, 24 sources were found to be located in groups or clusters of galaxies or to exhibit signs of jet reorientation. Four objects representing pairs of radio sources deserve special attention: the distance between their parent galaxies is only a few tens of kiloparsecs. 
Analysis of spectral indices revealed a decrease in the number of sources satisfying the condition $\alpha\leq-0.9$. According to data from before 1996 in the 365-3940 MHz range, 90 of the 113 sources met this criterion. 
According to new surveys, the number of such sources has decreased to 70 for the 340-3000 MHz band and to 39 for the 76-226 MHz band, according to GLEAM data. 
This trend can be explained by instrumental effects caused by differences in the angular resolution of the surveys, leading to errors in determining the total flux density, as well as by refinements in the low-frequency spectra, thanks to GLEAM data. 
For individual sources, the observed differences in spectral indices may be a consequence of evolutionary processes in the
spectrum or the source’s intrinsic variability.


\end{abstract}

\maketitle

\section{Introduction}
Radio galaxies with redshift $z > 2$ and luminosity at 500\,MHz L$_{500}>10^{27}$W$\cdotp$Hz$^{-1}$ are distinguished as a separate population of high-redshift radio galaxies (HzRGs)~\citep{2008A&ARv..15...67M}.

The components of the HzRG radiation are the emissions from dust, stars, and an active nucleus. Studies of the first two components show that HzRGs belong to the most massive stellar systems in the early Universe~\citep{2007ApJS..171..353S, 2009MNRAS.395.1099B, 2010ApJ...725...36D} and demonstrate signs of a massive galaxy at the stage of formation~\citep{2008A&ARv..15...67M}, as well as rapid accretion of matter onto a supermassive black hole (SMBH)~\citep{1997ApJS..109....1C, 2001A&A...366....7V, 2008A&A...491..407N, 2012A&A...548A..45D}.
The powerful submillimeter emission is also directly related to active star formation~\citep{2013MNRAS.429..744R}.
The dust torus of the radio galaxy obscures the light from the hot accretion disk and provides more opportunities to study the stellar population of the host galaxy than in the case of a quasar~\citep{2006ApJ...651..142H, 2010MNRAS.401.2531A}.

There is reason to believe that the galaxy and the SMBH are formed simultaneously~\citep{1998AJ....115.2285M, 2004ApJ...604L..89H, 2006ApJS..163...50H}.
According to the hierarchical model, the most massive star systems form at peaks of dark matter density by merging large numbers of small galaxies~\citep{1978MNRAS.183..341W}.
At $z\gtrsim 2\div2.5$, galaxy clusters are still in the process of forming, as there is not enough time for virialization.
For this reason, they are called protoclusters.
Observations show that HzRGs are most often found in fairly dense environments~\citep{2003Natur.425..264S, 2010MNRAS.405..347F, 2010MNRAS.405.2623S, 2012ApJ...749..169G, 2012A&A...539A..33M}, and protoclusters are likely to be found in their immediate vicinity.
Since HzRGs are found at high redshifts, they may mark galaxy clusters at cosmological distances.

Bright radio galaxies at $z>6$ can be used to study the reionization process in detail~\citep{2018MNRAS.475.5041S}. The red-shifted $\lambda = 21\ \text{cm}$ (1.4\,GHz) ultra-fine transition line of neutral hydrogen falls into the low-frequency radio range ($\nu<200$\,MHz) and can be observed as absorption in the spectra of the radio galaxy at $z>6$.
Such absorption lines can, in principle, be detected by current and next-generation radio telescopes~\citep{2002ApJ...577...22C, 2019PhDT.......148S}.

Blind searches for distant radio galaxies are ineffective. At high redshifts, identifying the host galaxy and determining its properties require extensive observation time and are often beyond the reach of most existing instruments.

\cite{1979A&AS...35..153T} discovered that distant radio sources have steep spectra. It was confirmed in subsequent studies, for example,~\cite{1990AJ.....99.1397K}. Although steep spectra are also observed in pulsars and dying radio galaxies, this criterion is often used in the selection of distant radio galaxy candidates. 
The effectiveness of this approach was demonstrated by~\cite{1994A&AS..108...79R, 1997A&A...326..505R, 2000A&AS..143..303D, 2004MNRAS.347..837D, 2006MNRAS.366...58D, 2007MNRAS.381..341B}. Additional criteria are also used: small angular sizes, a weak integrated flux density, the absence of candidates in the optical and infrared ranges, as well as the shape of the spectrum, which is a convex spectrum at low frequencies. Similar criteria were used \cite{2018MNRAS.480.2733S}, in which a radio galaxy with $z=5.72$ was found in their sample, currently the most distant known radio galaxy.

\subsection{The Big Trio program.}
In the SAO RAS, the Big Trio program~\citep {1992SvA....36..343G, 2000A&AT...19..297P} was launched in the early 1990s, aiming at searching for distant radio galaxies and their further study.
The selection of objects was made from radio sources discovered in the Cold Experiment\footnote{The Cold Experiment~\citep{1981PAZh....7..290B,1984SoSAO..41.....B,1984SoSAO..42.....B} consisted of a series of long-term surveys of a sky strip ($24^{h}\times 0.7^{\circ}$) at the declination of microquasar SS 433 ($\delta\approx5^{\circ}$),  carried out at several frequencies between 1980 and 1987. The aim of these surveys was to search for fluctuations in the cosmological background. The deepest flux density limit ($\simeq$2.5\,mJy) was reached at 3.94\,GHz.}
 and included in the RC (RATAN Cold) catalog~\citep{1991A&AS...87....1P, 1992A&AS...96..583P}.

The main selection criterion was the steepness of the radio spectrum ($\alpha\leq-0.9$). The two-frequency spectral indices of the sources were determined using data from RC (3.94\,GHz) and UTRAO (365\,MHz)~\citep{1980PAUTx..17....1D}.

For the SS (Steep Spectra) sample sources, the radio coordinates were refined, the morphological structure and angular sizes were determined using the observations available in the MIT–GB–VLA archive~\citep{1996ARep...40..759F}, and observations were also carried out on the VLA for other sources from the RC catalog~\citep{1995BSAO...40....5P}. 
In total, 389 radio maps were obtained for 208 RC sources at a frequency of 1.4 GHz with angular resolutions from 1.5$^{\prime\prime}$ to 4.5$^{\prime\prime}$. Some of the unresolved objects were observed at frequencies of 4.8, 8, and 14 GHz with a resolution up to 0.4$^{\prime\prime}$.

Within the framework of the Big Trio program in 1991-2003, 113 objects were observed with the 6-meter optical telescope BTA.
In addition, frames for 22 radio sources with subarcsecond visibility were obtained using the 2-meter NORDIC telescope~\citep{1999A&AS..134..505P}.
For most sources (94\%) in the SS sample, optical counterparts were found at frames with depth of $m_{R}\approx24.5^{m}$~\citep{1992AZh....69..673G,1995ARep...39..543K,1996BSAO...40....5P}.
Based on BVRI photometry data, photometric redshift estimates were made for 48\% of the sample~\citep{2002ARep...46..531V}.

Spectroscopic studies of hosts were carried out at the BTA with the SCORPIO focal reducer~\citep{2005AstL...31..194A}.
Spectra were obtained for 71 objects~\citep{1999ARep...43..275D, 2003ARep...47..377A, 2006AstL...32..433K, 2010ARep...54..675P}.
50\% of the hosts were classified as radio galaxies by the type of optical spectrum (narrow emission lines), and half of them have $z>1$.
A quarter of the hosts were classified as quasars (broad spectral lines), 70\% of which have $z>1$.
No lines were detected in the spectra of the remaining quarter of candidates due to the weak brightness of the objects in optics.

Among 54 radio sources with measured spectral redshifts, 5 galaxies and 5 quasars with $z>2$ were discovered, including two sources with $z>3$ and one with $z>4$. The last three sources with $z>3$ have extreme radio luminosities at 500\,MHz $L > 10^{28}\text{W}\cdotp\text{Hz}^{-1}$. Currently, RC\,J0311+0507, $z=4.514$~\citep{2006AstL...32..433K,2021A&A...654A..88W}, is one of the two most powerful radio galaxies known, along with 8C\,1435+635, with a luminosity of $L_{150}>10^{30}\text{W}\cdotp\text{Hz}^{-1}$~\citep{2018MNRAS.480.2733S}.

The SS sample of sources with steep spectra of the Big Trio program has been studied quite well. With the advent of deep sky surveys, it became possible to confirm or refine optical identifications, improve morphology, and radio spectra. 

The paper adopts the flat $\Lambda$CDM cosmology based on Planck's results: $H_0=67.4\ \text{km}\cdotp\text{s}^{-1}\cdotp\text{Mpc}^{-1}$, $\Omega_m=0.315$~\citep{2020A&A...641A...6P}.

The spectral index of the radio source $\alpha$ is defined as \mbox{$S_{\nu}\propto\nu^{\alpha}$}.

\section{Used information resources and software}
To study radio sources, we took information from the next surveys VLSSr~\citep{2014MNRAS.440..327L}, TGSS \citep{2017A&A...598A..78I}, GLEAM \citep{2017MNRAS.464.1146H}, VCSS \citep{2016ApJ...832...60P, 2021AAS...23721106P}, TXS \citep{1996AJ....111.1945D} and UTRAO~\citep{1980PAUTx..17....1D}, MRC~\citep{1991Obs...111...72L}, RACS~\citep{2021PASA...38...58H, 2024PASA...41....3D}, NVSS~\citep{1998AJ....115.1693C}, FIRST~\citep{2015ApJ...801...26H}, VLASS~\citep{2023ApJS..267...37G}, RC~\citep{1991A&AS...87....1P, 1992A&AS...96..583P}, PMN~\citep{1994ApJS...91..111W}, GB6~\citep{1996ApJS..103..427G}, PKS~\citep{1990PKS...C......0W}.

To clarify the morphological structure of radio sources, in addition to the existing collection of data obtained on the VLA for the Big Trio program~\citep{1996BSAO...40....5P}, we looked through radio maps from the GLEAM, TGSS, RACS, FIRST, VLASS surveys, and the NRAO archive.

To work with catalogs and surveys, the Aladin Sky Atlas software~\citep{2000A&AS..143...33B} was used, as well as to work with tables -- TOPCAT software~\citep{2005ASPC..347...29T}.

The radio spectra were plotted using the spg program from the FADPS data processing package~\citep{1993BSAO...36..132V} .

To search for information about the radio sources, we used the SIMBAD~\citep{2000A&AS..143....9W}, VizieR, CATS~\citep{2005BSAO...58..118V}, NED, DataLab~\citep{2020ASPC..522..153H} databases, as well as the optical surveys DecLS LS, DES~\citep{2018ApJS..239...18A, 2021ApJS..255...20A}, HSC-SPP~\citep{2019PASJ...71..114A, 2022PASJ...74..247A}, including near- and mid-infrared surveys LAS UKIDSS~\citep{2007MNRAS.379.1599L}, GPS UKIDSS~\citep{2008MNRAS.391..136L} and WISE~\citep{2012wise.rept....1C, 2021ApJS..253....8M}.

\section{Morphological structure of radio sources}
\begin{figure*}
\center
{(a)\includegraphics[scale=0.2]{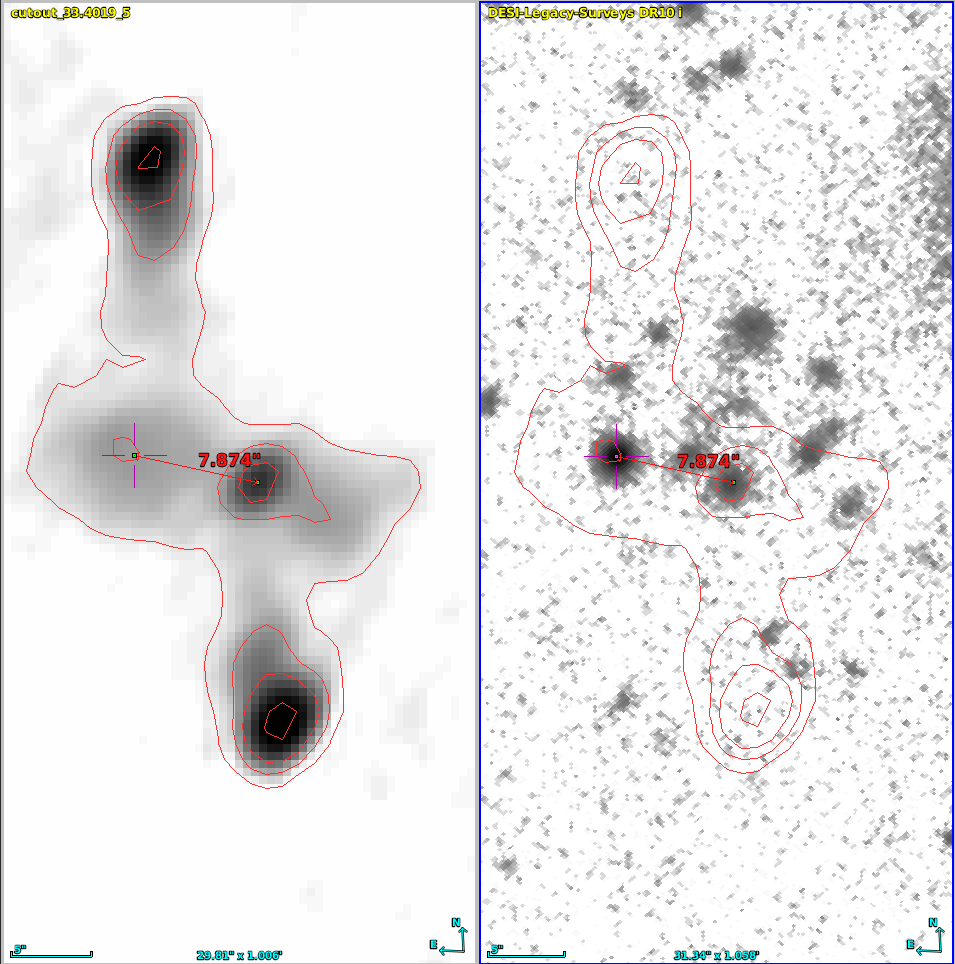}}
{(b)\includegraphics[scale=0.2]{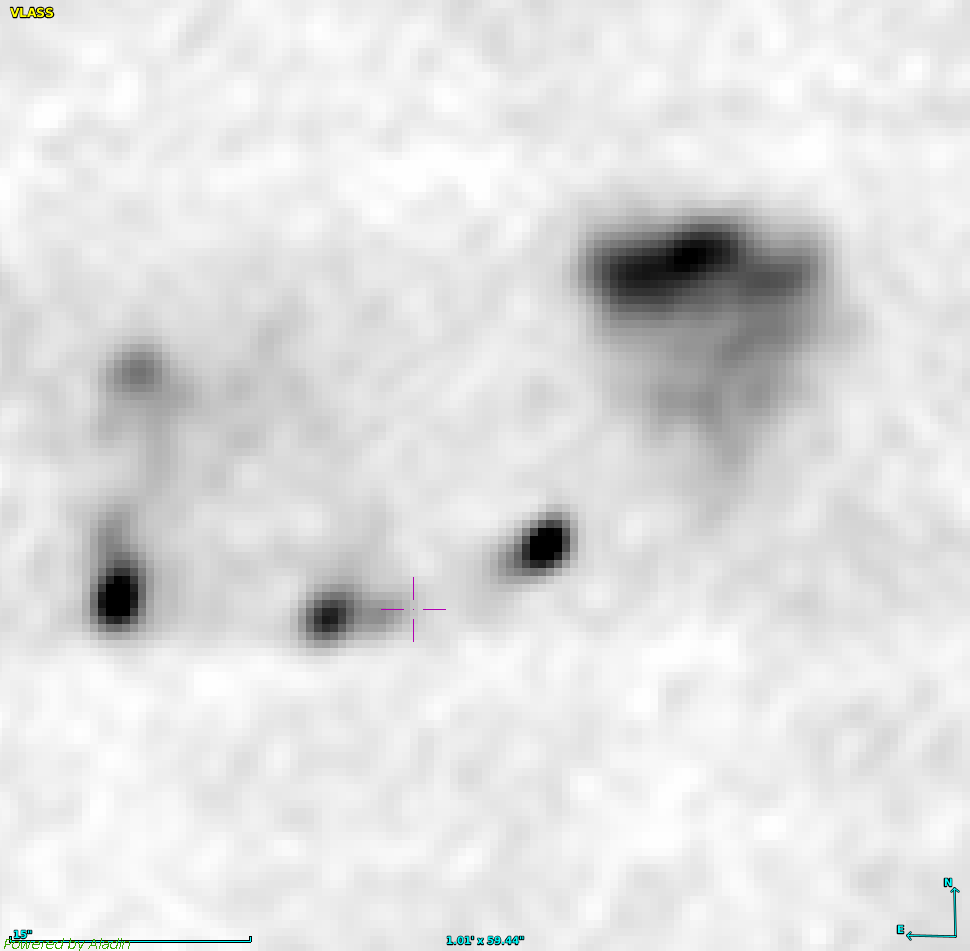}}
\caption{
Examples of radio sources: a) X-shaped radio source RC J0213+0516. Near its host galaxy ($z_{sp}$=0.935) there is a quasar ($z_{sp}$=0.934), associated with a weak radio source. The distance between the galaxy and the quasar is $7.9^{\prime\prime}$ or 61\, kpc; b) RC J0519+0510 is a DDRG-type radio source, which can also be classified as a rare morphological type of HSR sources.}
\label{fig:Pair}
\end{figure*}
For half of SS-sources, there are VLA observations with an angular resolution of $0.1^{\prime\prime}\div2.38^{\prime\prime}$~\citep{1995BSAO...40....5P, 1996ARep...40..759F}, and for the other half, there are maps from the VLASS survey with an angular resolution of $2.5^{\prime\prime}$.
For the radio source RC\,J0311+0507 (4C\,04.11), MERLIN and EVN maps with an angular resolution of $0.025^{\prime\prime}$~\citep{2014MNRAS.439.2314P} were obtained, and for RC\,J1740+0502, MERLIN maps with an angular resolution of $0.1^{\prime\prime}$.

According to these radio maps, point unresolved radio sources make up 18\% of the sample, with 9 of them being point sources on the VLASS maps and 11 unresolved on maps with higher angular resolution of $0.4^{\prime\prime}\div1.6^{\prime\prime}$.

For 44 double sources, the morphology was determined from the VLASS maps, and for 49 -- from maps with an angular resolution of $0.025^{\prime\prime}\div2.38^{\prime\prime}$.

Visual examination of radio maps and cutouts from the DESI, DES, HSC-SSP optical surveys, including the LAS UKIDSS and WISE infrared surveys, revealed that some radio sources are not single, but consist of two sources located close to each other on the picture plane. There were six such sources. Moreover, each of RC\,J0126+0502, RC\,J0213+0516, RC\,J0318+0506, and RC\,J1251+0446 forms a pair of close radio sources with angular distances between the parent galaxies of about $6^{\prime\prime}-8^{\prime\prime}$ or about 50--70 kpc. 

As an example, we consider RC\,J0213+0516 (Fig.~\ref{fig:Pair}, a), where the galaxy SDSS J021336.32+051819.0 is the host of a double radio source with the coordinates of the core $\alpha_{2000} = 02^{h}13^{m}36.32^{s}$ and $\delta_{2000} = +05^{\circ}18^{\prime}18.8^{\prime\prime}$ has a redshift of $z_{sp} = 0.935$~\citep{2010ARep...54..675P}. The quasar SDSS J021336.80+051820.7 has $z_{sp}^{SDSS} = 0.934$ and is associated with the weak radio source with coordinates $\alpha_{2000} = 02^{h}13^{m}36.77^{s}$ and $\delta_{2000} = +05^{\circ}18^{\prime}20.7^{\prime\prime}$. The angular distance between the hosts is $7.9^{\prime\prime}$ or 61\,kpc.

RC\,J0318+0456 was considered a double radio source based on NVSS and TGSS maps. Higher-resolution VLASS maps show that each component is an independent radio source. This is confirmed by identifying the parent objects for each component. The same is true for RC\,J0324+0442, where the components also appear to be independent sources.

Among the double radio sources, there are sources of a non-standard structure, which cannot be classified as FRI or FRII types~\citep{1974MNRAS.167P..31F}; that is, they are classified as FRI/FRII-type or so-called hybrid radio sources HyMoRS (HYbrid MOrphology Radio Sources)~\citep{2000A&A...363..507G, 2017AJ....154..253K,2022ApJ...941..136S}. 
Double radio sources with double lobes DDRG (Double-double Radio Galaxy)~\citep{2000MNRAS.315..371S, 2006MNRAS.366.1391S} also do not quite fit into the FRII-type.
In the SS sample, we assigned 8 radio sources to the hybrid type and 3 radio sources to the DDRG type. Example of double-double radio source  RC\,J0519+0510 is shown in Fig.~\ref{fig:Pair} (b).

A third (27\%) of the sample have a core, and in 16 of them the contribution of the core to the integrated flux density at 3\,GHz is less than 5\%, and in 15 double sources with a core -- from 10\% to 60\%. We assigned double sources with a core whose contribution to the integrated flux density at 3\,GHz is more than 10\% to triple sources. In addition, for 6 radio sources the contribution of the core is more than 35\%, and they can be classified as CDT (Core-Dominated Triple)~\citep{2006A&A...448..479M}.

There are radio sources with deformed lobes. They are classified by the following types -- winged sources WRG (Winged Radio Galaxy)\footnote{WRG include X-, S-, Z-shaped radio galaxies (XRG, ZRG).}~\citep{2007AJ....134.1245C, 2019ApJS..245...17Y, 2022ApJS..260....7B}, ``head-tail'' (HT) radio sources\footnote{HT galaxies include WAT (Wide-Angle Tailed), NAT (Narrow-Angle Tailed) and C-shaped radio galaxies.}\citep{1976ApJ...203L.107R, 2000ApJ...531..118B,2019A&A...626A...8M, 2022ApJS..259...31S}, as well as recently discovered radio sources with a ring-shaped diffuse radio emission -- ORC (Odd Radio Circle)~\citep{2021Galax...9...83N} and HSR (Horseshoe-Shaped Ring)~\citep{2024A&A...683A.175K}.

In the SS sample, we classified 12 WRGs, 8 WAT sources, and two more sources that are C-shaped or even horseshoe-shaped (HSR). RCJ0519+0510 is shown in Figure ~\ref{fig:Pair}(b) as such an example. 

The sample included two giant radio galaxies~\citep{1974Natur.250..625W,2001A&A...370..409L,2021Galax...9...99A} -- RC\,J0152+0453 and RC\,J1333+0451 with projected linear sizes of 990 and 1100 kpc, respectively, as well as twenty-six compact radio sources with sizes from 0.7 to 19 kpc or with an angular size less than $3^{\prime\prime}$, if the redshift is not known for the host of the unresolved radio source.

\section{Analysis of total flux densities and spectral indices}

Using a representative set of flux density measurements, we calculated spectral indices for each source in the SS sample. 
The calculations employed a linear fit to the flux densities across three distinct datasets, which covered the following frequency ranges: 365–3940 MHz (data prior to 1996), 340–3000 MHz (from recent catalogs), and 76–227 MHz (GLEAM survey data). 
A comparative analysis of the results revealed a systematic decrease in the proportion of sources with spectral indices satisfying the condition $\alpha\leq-0.9$.

When performing linear fits to flux densities, the endpoints of the spectral coverage can have a significant impact on the resulting spectral indices, particularly when the number of data points is limited. To address this issue, we conducted a targeted comparison of old and new measurements at specific frequencies: 365 MHz (TXS) versus 340 MHz (VSSS), as well as 1400 MHz (NVSS) versus 1368 MHz (RACS). According to established astrophysical concepts, sources with steep spectra typically exhibit greater temporal stability than those with flat spectra.

To identify sources with significant discrepancies between measurements, we implemented a two‑step procedure. First, we calculated the variability index for each source. Second, we performed a detailed analysis of sources with variability indices exceeding 3. This threshold enables the selection of objects exhibiting significant discrepancies among the data points.

Particular attention was paid to the data at 340 MHz and 365 MHz, as these frequencies are critical for determining spectral steepness. Refining the flux densities at these frequencies allows for a more accurate interpretation of the observed decrease in the number of sources with $\alpha\leq-0.9$. This analysis is essential for distinguishing instrumental effects from intrinsic source properties.

\subsection{Variability indices}

For the vast majority of sources in the sample, sufficient data are available to compare total flux densities across two or more epochs at different frequencies. We performed comparisons at 340--365~MHz (VCSS, TXS), 1368--1400~MHz (RACS-mid, NVSS), 3900--3940~MHz (GAISh, Cold), and 4850--5000~MHz (MIT-GB, GB6, PMN, PKS).

The time interval between the observations of the VCSS and TXS surveys is approximately 40 years, assuming the observation epoch is defined as the mean of the survey's start and end years.  
The interval between the mean epochs of the RACS-mid and NVSS surveys is 27 years.  
For individual sources observed in both the Zelenchuk-GAISh survey~\citep{1985SoSAO..47....5A} and the Kholod series of surveys, the time interval ranges from 2 to 19 years.  
In the MIT-GB~\citep{1986ApJS...61....1B}, GB6~\citep{1996ApJS..103..427G}, PMN~\citep{1995ApJS...97..347G}, and PKS~\citep{1975AuJPA..34...63S, 1981AuJPh..34..407B} surveys, the interval between observations can be as long as 18 years.

Note that data coverage is as follows: 340-365\,MHz -- 77\% of the sample; 1368--1400\,MHz -- 100\%; 3900-3940\,MHz -- 89\%; and 4850-5000\,MHz -- 82\%.

We used a simple variability criterion:
\[
V = \frac{S_{\max} - S_{\min}}{\sqrt{\sigma^2_{\max} + \sigma^2_{\min}}} > 3,
\]
where $S_{\max}$ and $S_{\min}$ are the maximum and minimum flux densities, and $\sigma_{\max}$ and $\sigma_{\min}$ are their respective measurement uncertainties.
\begin{figure}
\center
{(a)\includegraphics[scale=0.23]{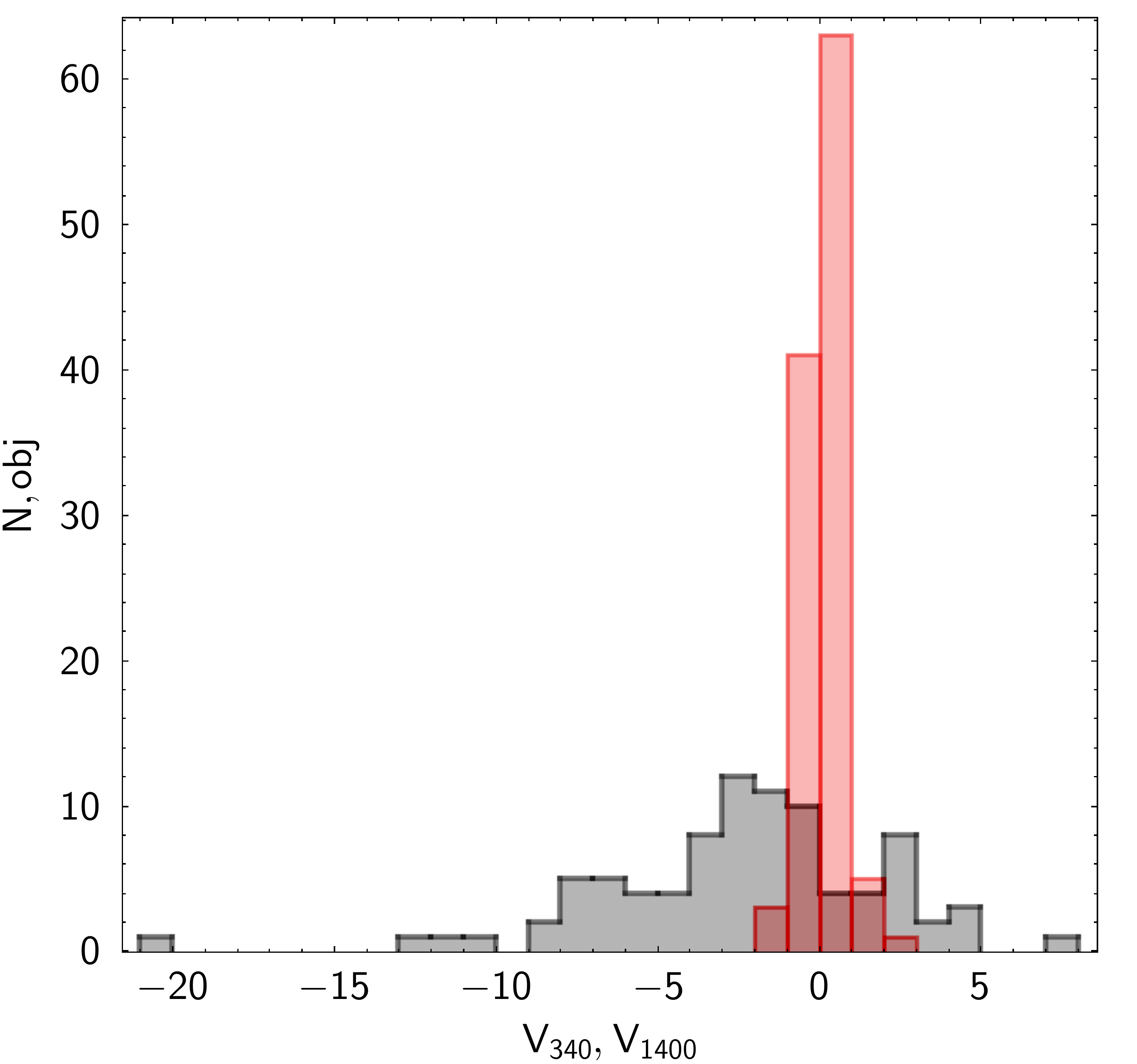}}
{(b)\includegraphics[scale=0.23]{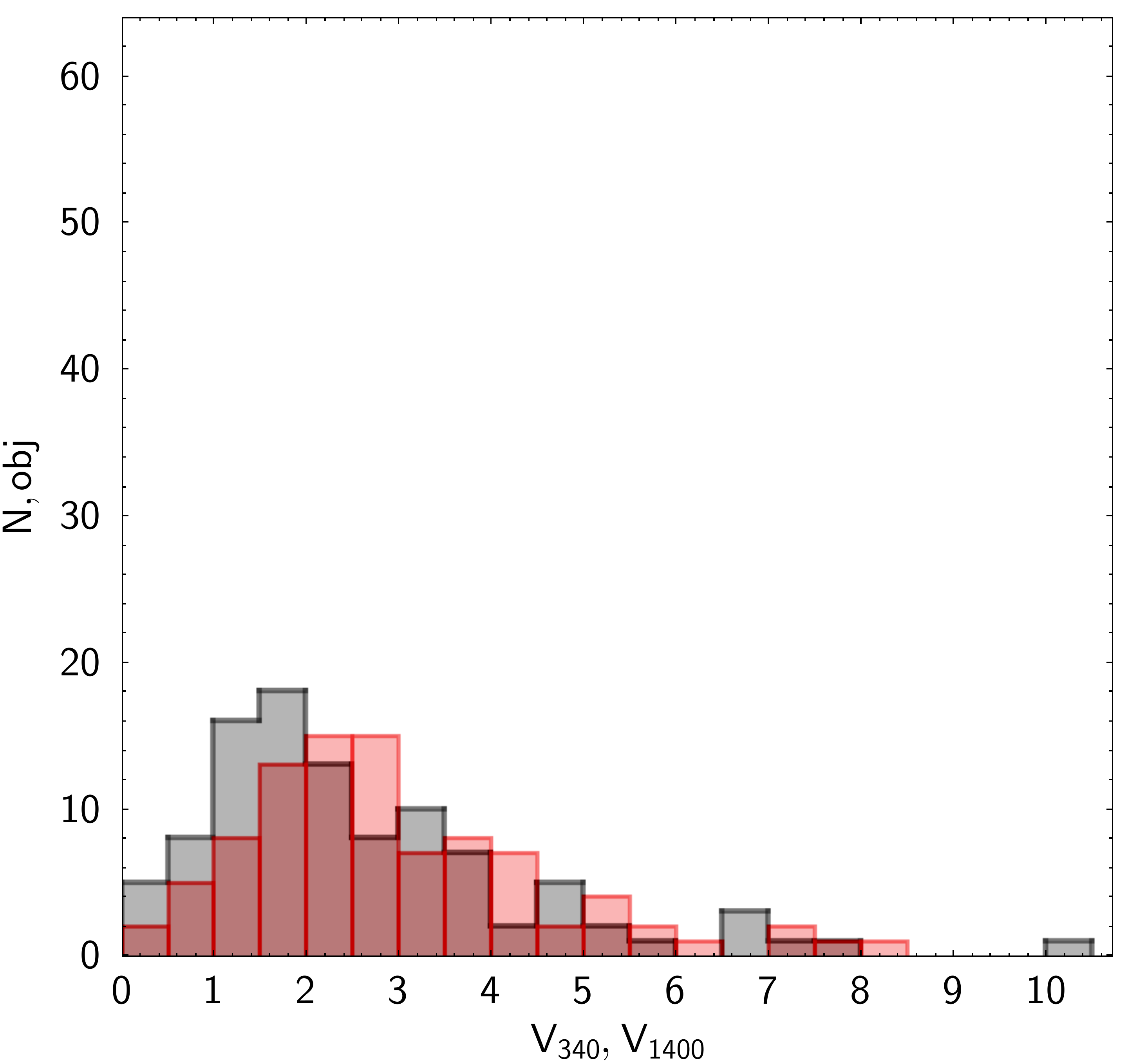}}
\caption{
Histograms of variability index distributions: (a) $V_{340}$ (grey) and $V_{1400}$ (red); (b) $V_{3940}$ (grey) and $V_{4850}$ (red).
}
\label{fig:Var}
\end{figure}

This criterion was used to calculate the variability indices $V_{3940}$ for the \mbox{3900--3940\,MHz} range and $V_{4850}$ for the \mbox{4775--5000\,MHz} range.

The variability index $V_{340}$ for the frequencies 340 and 365\,MHz, based on VCSS and TXS data, as well as the index $V_{1400}$ for 1368 and 1400\,MHz, based on RACS-mid and NVSS data, were determined as follows:
\[
V_{340} = \frac{S_{340} - S_{365}}{\sqrt{\sigma^2_{340} + \sigma^2_{365}}}, \quad
V_{1400} = \frac{S_{1368} - S_{1400}}{\sqrt{\sigma^2_{1368} + \sigma^2_{1400}}},
\]
where $S_{340}$ and $S_{365}$ are the flux densities at 340 and 365\,MHz, respectively, and $\sigma_{340}$ and $\sigma_{365}$ are their measurement uncertainties. Similarly, $S_{1368}$ and $S_{1400}$ denote the flux densities at 1368 and 1400\,MHz, with corresponding uncertainties $\sigma_{1368}$ and $\sigma_{1400}$.

This formulation allows us to examine whether $S_{340}$ exceeds $S_{365}$ or vice versa. As a result, the indices $V_{340}$ and $V_{1400}$ can take negative values.

Histograms showing the distributions of $V_{340}$ and $V_{1400}$ are presented in Figure~\ref{fig:Var}(a), while those for $V_{3940}$ and $V_{4850}$ are shown in Figure~\ref{fig:Var}(b).

A notable and unexpected difference was observed between the distributions of $V_{340}$ and $V_{1400}$. All 113 radio sources in the sample exhibit variability indices at 1368-1400\,MHz satisfying $|V_{1400}| < 3$. In contrast, among the 87 sources with available data in the VCSS and TXS catalogs, 44\% show $|V_{340}| > 3$.

At frequencies of 3900--3940\,MHz, 31\% of the 101 radio sources have $V_{3940} > 3$, while at 4850--5000\,MHz, 37\% of the 93 sources exhibit $V_{4850} > 3$.

For 47 radio sources, at least one of the indices $V_{3940}$ or $V_{4850}$ exceeds 3; for 10 sources, both indices are greater than 3.  
Notably, for RC\,J0934+0505, RC\,J1142+0455, RC\,J1456+0456, and RC\,J2225+0523?all three indices, $|V_{340}|$, $V_{3940}$, and $V_{4850}$, are greater than 3.

Next, we examine the reasons behind the pronounced difference in the variability index $V_{340}$ compared to $V_{1400}$.

\subsection{Flux density measurement errors}

We inspected the vicinity of the radio sources using cutouts from the GLEAM, TGSS, RACS-low, and RACS-mid surveys. These cutouts were examined for nearby radio sources that could contribute additional flux to $S_{365}$, as well as for other factors that might affect the value of $S_{340}$.

\subsubsection{Contribution of nearby radio sources}

As noted by \citet{1996AJ....111.1945D}, two sources of comparable strength are listed separately in the TXS catalog if they are separated by more than $9.6^{\prime}$ in right ascension ($\alpha$) or more than $8.1^{\prime}$ in declination ($\delta$).

To account for the contribution of nearby radio sources to $S_{365}$ for the target object, we adopted a search region of approximately the same size. Using RACS-low cutouts, we identified sources within this region, summed their flux densities, and calculated their total fraction relative to the flux of the target source. This fraction was then subtracted from the integrated flux density $S_{365}$ to obtain the corrected value $S_{365}^{c}$.

For 16 out of 87 radio sources, no neighboring sources were detected within the examined regions.

\begin{figure}
\center
\includegraphics[scale=0.23]{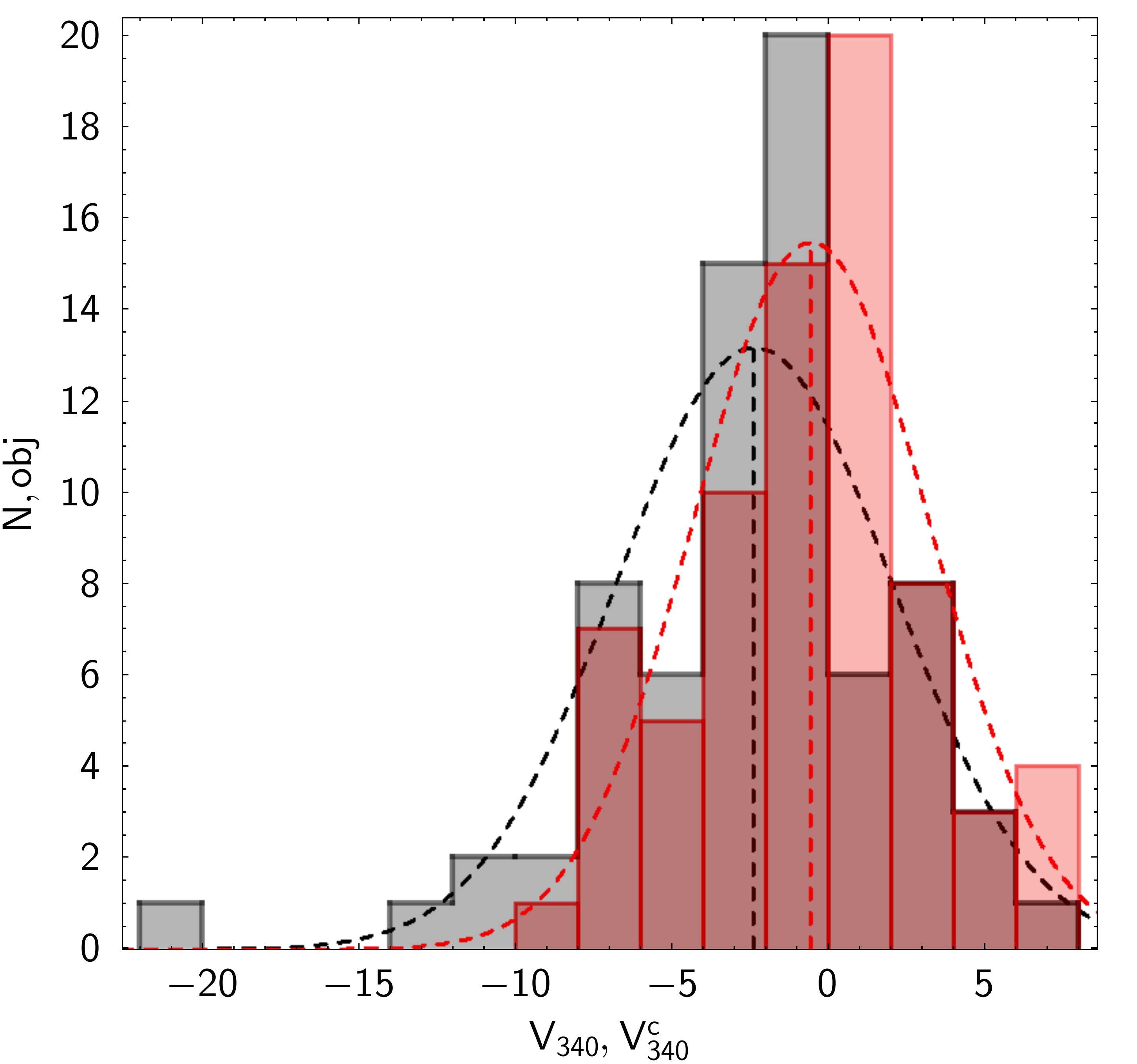}
\includegraphics[scale=0.23]{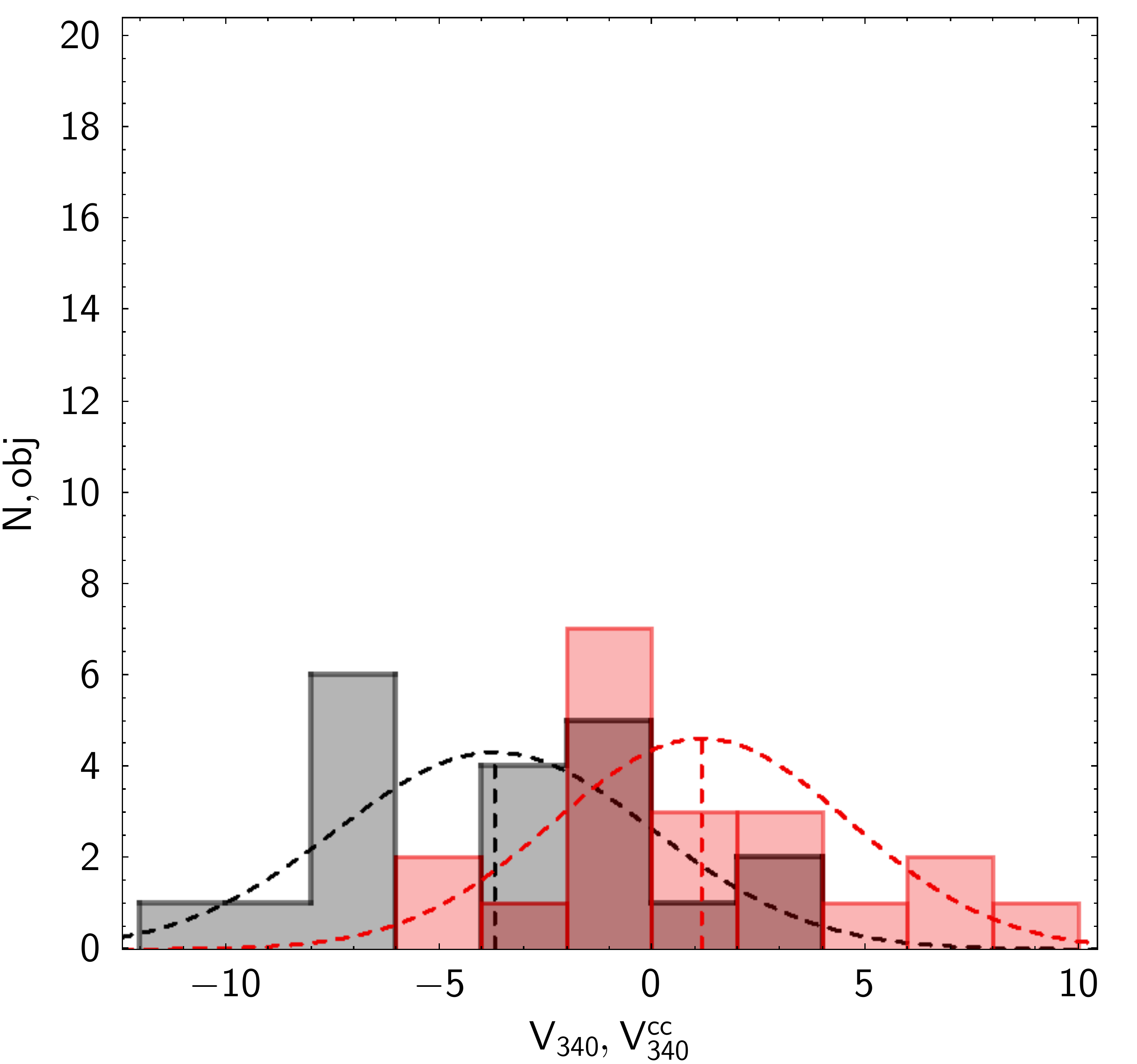}
\caption{
Distributions of variability indices: (a) $V_{340}$ (gray) for sources with nearby companions, and $V_{340}^{c}$ (red), where the contribution of neighbors is accounted for; (b) $V_{340}$ (gray) and $V_{340}^{cc}$ (red) for double sources. The index $V_{340}^{cc}$ incorporates both the underestimated flux $S_{340}$ of double radio sources and the contribution of neighbors. Gaussian fits to the distributions are shown as dashed lines in black and red, respectively.
}
\label{fig:Var1}
\end{figure}

Figure~\ref{fig:Var1}(a) presents the distributions of the variability indices $V_{340}$ (gray) and $V_{340}^{c}$ (red) for sources with nearby companions. The corrected index $V_{340}^{c}$ accounts for the fractional contribution of neighboring sources to the total flux density $S_{365}$.

The mean value of the variability index was $V_{340} = -2.38$, and after correcting for the contribution of neighbors, it became $V_{340}^{c} = -0.56$.

\subsubsection{Errors in total flux densities for double radio sources}

From the VCSS catalog, we selected double radio sources with angular sizes $LAS > 15^{\prime\prime}$, exceeding the angular resolution of the VCSS survey. The largest angular size (LAS) was measured between the outermost VLASS components of each radio source.  
The sample includes 28 double sources with angular sizes ranging from $15^{\prime\prime}$ to $124^{\prime\prime}$.

It was found that the integrated flux density of some sources in the VCSS catalog is not always accurately determined. This issue is particularly evident for sources with asymmetric flux density components, as seen in the FIRST and VLASS survey maps. Such sources may be represented in VCSS by a single component, leading to an underestimated total flux density $S_{340}$.

Notably, only two out of the 28 double radio sources RC\,J0318+0456 ($LAS = 77.7^{\prime\prime}$) and RC\,J1148+0455 ($LAS = 40.5^{\prime\prime}$) are listed with two components in the VCSS catalog.

For 11 double radio sources with $LAS > 15^{\prime\prime}$, corrections to the integrated flux density $S_{340}$ were made using data from other radio catalogs. As a result, the corrected flux density $S_{340}^{cc}$ increased by 7\%--50\% relative to the original $S_{340}$. Consequently, their variability indices decreased in absolute value and fell below the threshold of 3.

Figure~\ref{fig:Var1}(b) shows histograms of the variability index distributions: $V_{340}$ (gray) and $V_{340}^{cc}$ (red) for the double radio sources. In calculating $V_{340}^{cc}$, both the contribution of nearby sources to $S_{365}$ and the correction of $S_{340}$ for double sources were taken into account.

The average value of the variability index for double sources was $V_{340} = -3.70$. After correcting for the underestimation of the integrated flux densities, the revised value became $V_{340}^{cc} = 1.15$.

Nevertheless, even after these corrections, the variability index for the double radio sources RC\,J1031+0443, RC\,J1142+0455, RC\,J2029+0456, and RC\,J2036+0451 remained at the level of $|V_{340}^{cc}| > 3$.

We suggest that the variability index $V_{340}$ for these sources is likely overestimated, as the correction of the underestimated flux density $S_{340}$ was unsuccessful.

\begin{table*}
\caption{Parameters of radio sources with a significant variability index.\\
$Name$ -- radio source; $V_{340}^{c}$, $V_{3940}$, $V_{4850}$ -- variability indices at 340, 3940 and 4850 MHz; -- $Type$ -- the type of the host: Q -- quasar, G -- galaxy,  EF -- empty field; $mag_{r}$ -- the magnitude in the r-filter; $LAS$ -- the angular size; $RType$ -- the type of radio source: P -- point, D -- double, T -- triple, cdt -- triple with dominant core, css -- compact radio source with steep spectrum; $z$ -- the redshift; $mz$ -- ``s'' -- spectroscopic z, ``p'' -- photometric z. The ``?'' character indicates an undefined type definition; the ``*'' character marks a presumably variable source.}
\centering
\begin{tabular}{|l|c|c|c|c|c|c|c|c|c|}
\hline
~~~~~$Name$ &~$V_{340}^{c}$~&~$V_{3940}$~&~$V_{4850}$~&~$Type$~&~$mag_{r}$~~&~$LAS$,~        &~$Rtype$~&~~$z$ &~$mz$~ \\
          &               &            &            &        &           & $^{\prime\prime}$ &          &     & \\
\hline \hline
\multicolumn{10}{|l|}{Peaked-spectrum radio sources} \\
\hline
1. RC\,J0133+0459~&  2.7 & 0.1 & 1.7 & G? & 25.1  & 0.2 & P     &  & \\
2. RC\,J0250+0512 & -9.8 & 2.6 & 3.9 & EF & $>$25 & 1.2 & D/css &  & \\
3. RC\,J0907+0439 & -5.7 & 5.0 & 1.8 & G? & 24.7  & 0.3 & P     &  & \\
4. RC\,J1100+0444* & -3.1 & 4.8 & 1.8 & Q & 19.1  & 0.3 & P     & 0.886 & s\\
5. RC\,J1150+0459 & -4.8 & 1.7 & 2.6 &  G?  & 23.3 & 0.3 & P     & 1.27 & s \\
\hline \hline
\multicolumn{10}{|l|}{Radio sources with $V_{340}^{c}>3$} \\
\hline
1. RC\,J0135+0450* & 3.5 & - & 1.5 & Q? & 18.6 &  7.8 & T/cdt & 0.372 & s\\
2. RC\,J0143+0505* & 4.1 & - & 4.2 & Q & 20.8 &  7.4 & T     & 2.135 & s\\
3. RC\,J1456+0456* & 7.2 & 3.5 & 3.1 & Q & 20.1 &  2.2 & P     & 2.136 & s\\
4. RC\,J1503+0456* & 4.9 & 1.7 & 1.2 & Q? & 22.9 &  4.5 & T/cdt & 0.788 & s\\
5. RC\,J1551+0458* & 3.1 & 1.0 & 1.2 & G & 23.9 & 11.6 & D     & 1.29 & p\\
6. RC\,J1703+0502 & 4.1 & 0.8 & 2.8 & G? & 23.8 &  1.8 & D/css & 1.23 & p\\
\hline \hline
\multicolumn{10}{|l|}{Radio sources with $V_{340}^{c}<-3$} \\
\hline
1. RC\,J0034+0513 & -6.5 & 1.5 & 1.8 & G & 23.1 & 12.1 & T?    & 0.962 & s \\
2. RC\,J0226+0512* & -7.6 & 2.6 & 1.9 & Q & 20.1 & 10.7 & D;dd  & 1.242 & s \\
3. RC\,J0311+0507 & -5.8 & 1.3 & 2.9 &  G  & 22.9 &  2.8 & D;mc  & 4.508 & s \\
4. RC\,J0355+0449 & -4.8 & 1.4 & 7.7 &  G? & 24.1 &  2.4 & D;css & 2.7   & p \\
5. RC\,J0934+0505 & -5.2 & 3.1 & 7.0 &  G? & 23.0 &  5.0 & D     & 1.68  & p \\
6. RC\,J1011+0502 & -4.0 & 1.6 & 2.1 &  G? & 23.8 &  2.8 & D;css &       &   \\
7. RC\,J1124+0456* & -7.6 & 2.7 & 3.7 & G  & 17.3 & 11.9 & D;mc  & 0.284 & s \\
8. RC\,J1154+0431* & -3.6 & 1.2 & 2.0 &  Q  & 19.3 &  6.7 & D     & 0.988 & s \\
9. RC\,J2225+0523* & -3.8 & 4.9 & 4.4 &  Q  & 17.8 &  2.7 & T     & 2.323 & s \\
\hline
\end{tabular}
\label{tab:RCvar}
\end{table*}
We suggest two possible reasons for the underestimation of the integrated flux density of $S_{340}$.
The first is related to the processing of double sources with $LAS > 15^{\prime\prime}$ in the VCSS survey, which may lead to an incomplete flux reconstruction.
The second reason is the presence of extended regions of low-surface-brightness emission, which are not detected by VCSS but may contribute significantly to the total flux density in lower-resolution surveys such as GLEAM and TXS.

\subsection{Radio sources with $|V_{340}^{c}>3|$}
\begin{figure*}
\center
{a)\includegraphics[scale=0.165]{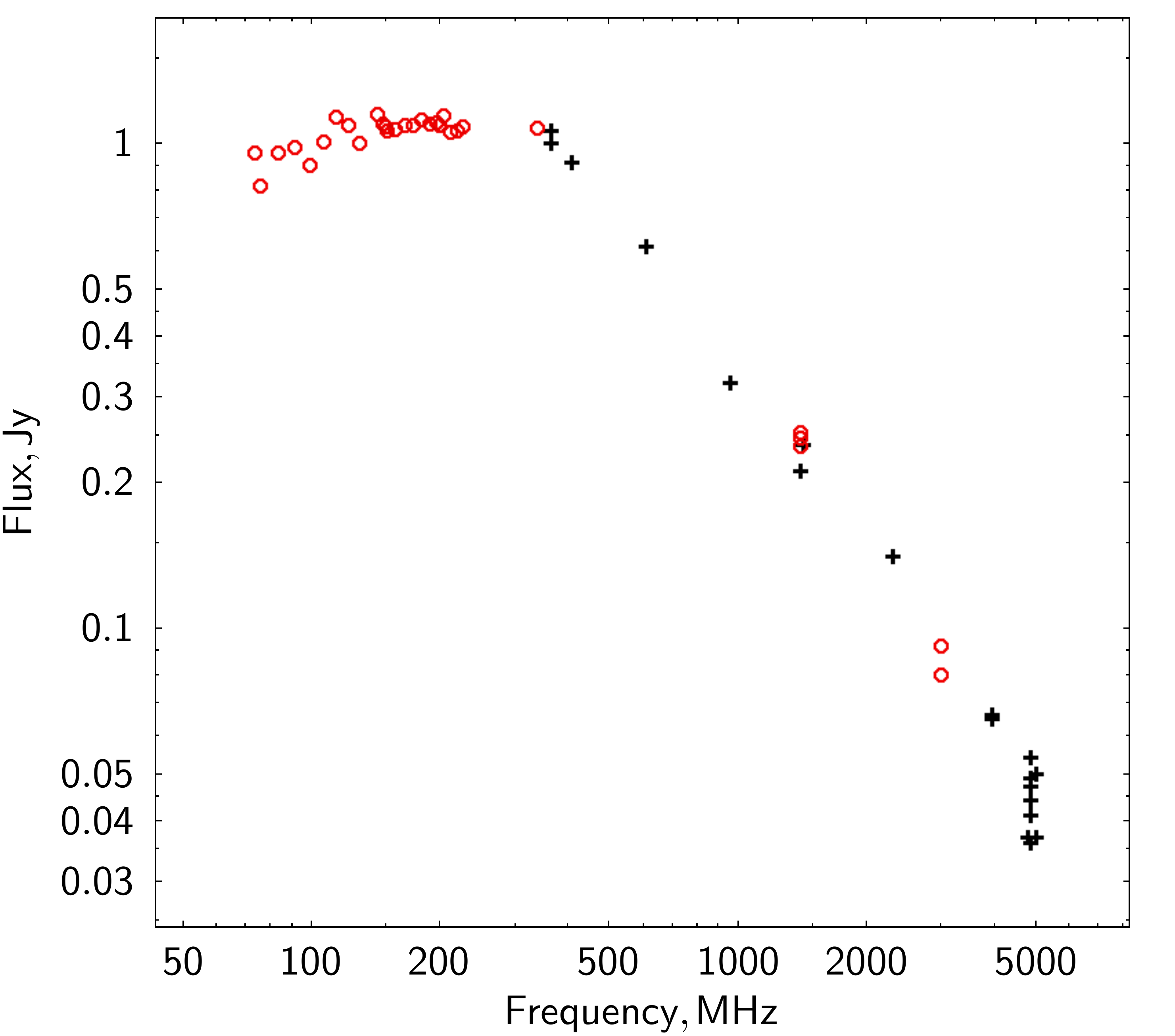}}
{b)\includegraphics[scale=0.165]{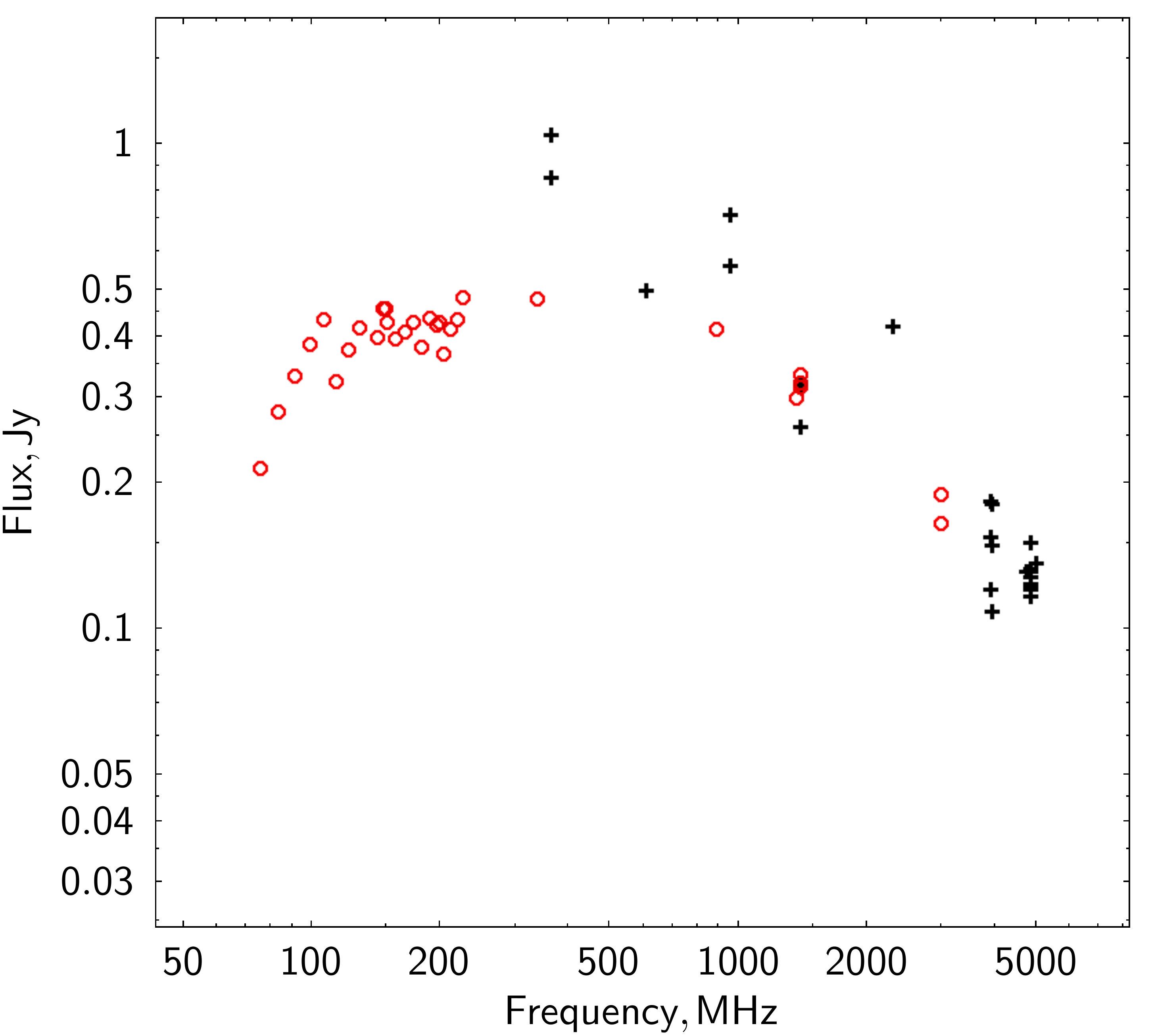}}
{c)\includegraphics[scale=0.165]{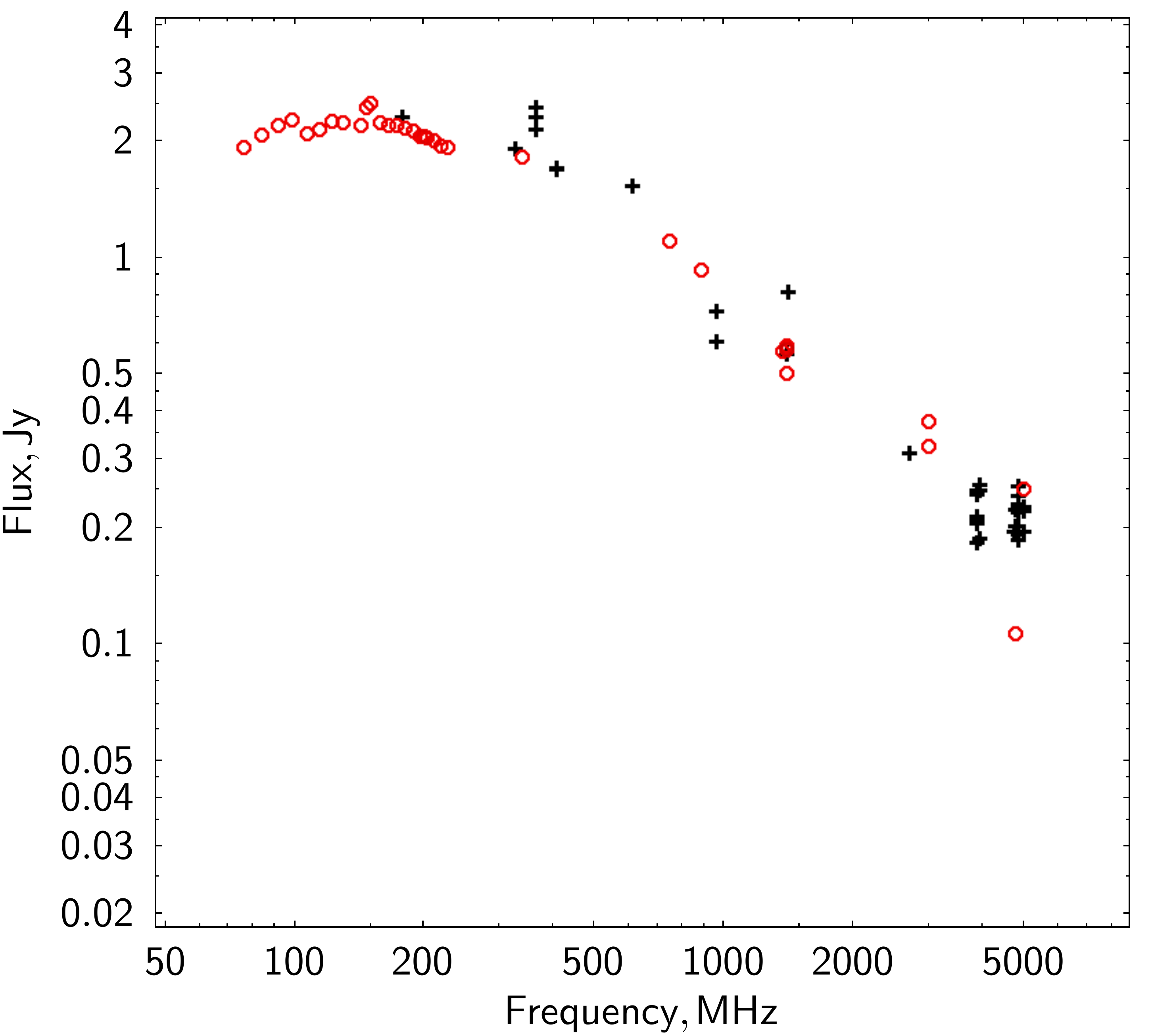}}
\caption{
Radio spectra of sources exhibiting a spectral peak:  
(a) RC\,J0133+0459, (b) RC\,J0907+0439, (c) RC\,J1100+0444.  
The spectra were constructed using archival data from~\citet{1996BSAO...42....5B} and newly obtained measurements from the GLEAM, VCSS, RACS, and VLASS surveys (marked in red).
}
\label{fig:SpPS}
\end{figure*}
If we exclude sources for which the contribution to the flux density $S_{365}$ from nearby companions has been accounted for, 19 sources remain with $|V_{340}^{c}| > 3$.  
The parameters of these sources are listed in Table~\ref{tab:RCvar}.  
The type and optical magnitude of the host object were determined from the LS survey, while redshift values were obtained from \citet{2010ARep...54..675P}, SDSS, and LS.

Table~\ref{tab:RCvar} presents the parameters for three groups of radio sources:  
five sources with a spectral peak in the 100--500\,MHz range,  
six sources with $V_{340}^{c} > 3$,  
and nine sources with $V_{340}^{c} < -3$.

Eight sources in the SS sample exhibit a spectral peak at frequencies between 100 and 500\,MHz.  
Five of these have data available in the VCSS survey.  
Among them, four sources show $V_{340}^{c} < -3$, while RC\,J0133+0459 has a variability index of $V_{340}^{c} < 3$.

All five sources have small angular sizes (see Table~\ref{tab:RCvar}) and are unresolved in the VLASS survey.  
Their host objects are faint galaxies with apparent magnitudes in the range of $23^{m}-25^{m}$ in the optical, except for the quasar RC\,J1100+0444.

\citet{2021AN....342.1151O} noted that during the adiabatic expansion of homogeneous synchrotron sources, a shift of the spectral peak toward lower frequencies can be observed in the continuum spectrum.

The radio spectra of the MPS sources RC\,J0133+0459, RC\,J0907+0439, and RC\,J1100+0444, are shown in Fig.~\ref{fig:SpPS}. Among these, no spectral shift is observed for RC\,J0133+0459 (Fig.~\ref{fig:SpPS}a), while RC\,J0907+0439 exhibits the most pronounced shift (Fig.~\ref{fig:SpPS}b). The quasar RC\,J1100+0444 (Fig.~\ref{fig:SpPS}c) is a variable source that may display a spectral peak during periods of activity.

We propose that the lower flux density $S_{340}$ compared to $S_{365}$ for these MPS sources, with the exception of RC\,J1100+0444, is due to a shift in the spectral peak caused by radio source expansion. This effect is evident in the TXS and VCSS data, which are separated by an interval of 34–48 years between observing epochs.

Nine sources with $V_{340}^{c} < -3$ (see Table~\ref{tab:RCvar}) have small angular sizes ($LAS < 8^{\prime\prime}$). Four of them are associated with quasars and are most likely variable objects. The remaining five sources are even more compact and are presumably young, exhibiting a low-frequency spectral shift similar to that seen in MPS sources. Including MPS sources without VCSS data, this group comprises 13 sources (12\% of the sample) that are likely young.

Thus, the presence of a large proportion of sources with $|V_{340}| > 3$ in the sample is primarily explained by differences in the angular resolution of the using surveys, the contribution of nearby sources to the total flux density $S_{365}$, and the underestimated values of $S_{340}$ for some double sources.

Ultimately, we conclude that sources with $|V_{340}| > 3$ may include young, rapidly evolving objects, variable sources, and sources with extended low-surface-brightness components, likely formed during a previous episode of radio activity. 
Such extended components may not be detectable in the VCSS survey, but are more likely to be observed in TXS and GLEAM due to their lower angular resolution and, consequently, higher sensitivity to extended low-surface-brightness components.

Table~\ref{tab:RCvar} lists ten sources: RC\,J0135+0450, RC\,J0143+0505, RC\,J0226+0512, RC\,J1100+0444, RC\,J1124+0456 (4C+05.50), RC\,J1154+0431, RC\,J1456+0456, RC\,J1503+0456, RC\,J1551+0458, and RC\,J2225+0523 that we classify as variable based on the available data. Notably, in the Cold experiment surveys, the sources RC\,J0506+0508, RC\,J1124+0456, RC\,J1213+0500, and RC\,J1551+0458 were also identified as variable radio sources~\citep{2012AstBu..67..318M, 2015AstBu..70...33M}.

In total, 12 radio sources (11\% of the sample) exhibit variability in their integrated flux density.

\section{Spectral indices of sources of the SS-sample}
The initial selection of candidates for the sample of steep-spectrum sources studied in the Big Trio program was based on the criterion $\alpha_{365}^{3940} \leq -0.9$. Of the 88 sources mentioned by
\cite{1994AZh....71..684S} and included in the program, 83\%, or 94\%, met this criterion.

We compared the spectral indices of the SS sample sources, calculated from spectra constructed by fitting the following sets of flux densities:
\begin{enumerate}
    \item[\textbf{old:}] 
    The spectral index $\alpha_{365}^{3940}$ between 365 and 3940\,MHz was determined by linear approximation using data from~\citet{1996BSAO...42....5B}, which provides the most comprehensive collection of measurements obtained before 1996, as well as RATAN-600 observations~\citep{1991A&AS...87....1P, 1992A&AS...96..583P, 1994AZh....71..684S, 1996BSAO...40..128B, 1996BSAO...42....5B, 1996BSAO...40....5P, 2010AstBu..65...42S, 2017AstBu..72..150Z, 2018AstBu..73..142Z}.    
    \item[\textbf{new:}] 
    The spectral index $\alpha_{340}^{3000}$ was calculated over the frequency range 340-3000\,MHz using a linear fit to data from VCSS, RACS-low, RACS-mid, NVSS, VLASS, and additional catalogs~\citep{2017AJ....154..156T, 2018MNRAS.474.5008D, 2021ApJ...914...42B, 2023ApJS..267...37G}. This range was chosen to approximately match the 365-3940\,MHz interval.
    \item[\textbf{glm:}] 
    The spectral index $\alpha_{76}^{227}$ was derived by linear fitting of GLEAM data.
    \item[\textbf{all:}] 
    Spectral indices $\alpha_{74}$ and $\alpha_{7700}$ were obtained by parabolic fitting of all available flux density measurements for each source at 74 and 7700\,MHz, respectively.  
    The spectral curvature parameter was calculated as
    $SCP = \alpha_{74} - \alpha_{7700}$,
    which is commonly used to estimate the evolutionary stage of a radio source~\citep{2011A&A...526A.148M}.
\end{enumerate}
\begin{figure}
\center
{a)\includegraphics[scale=0.23]{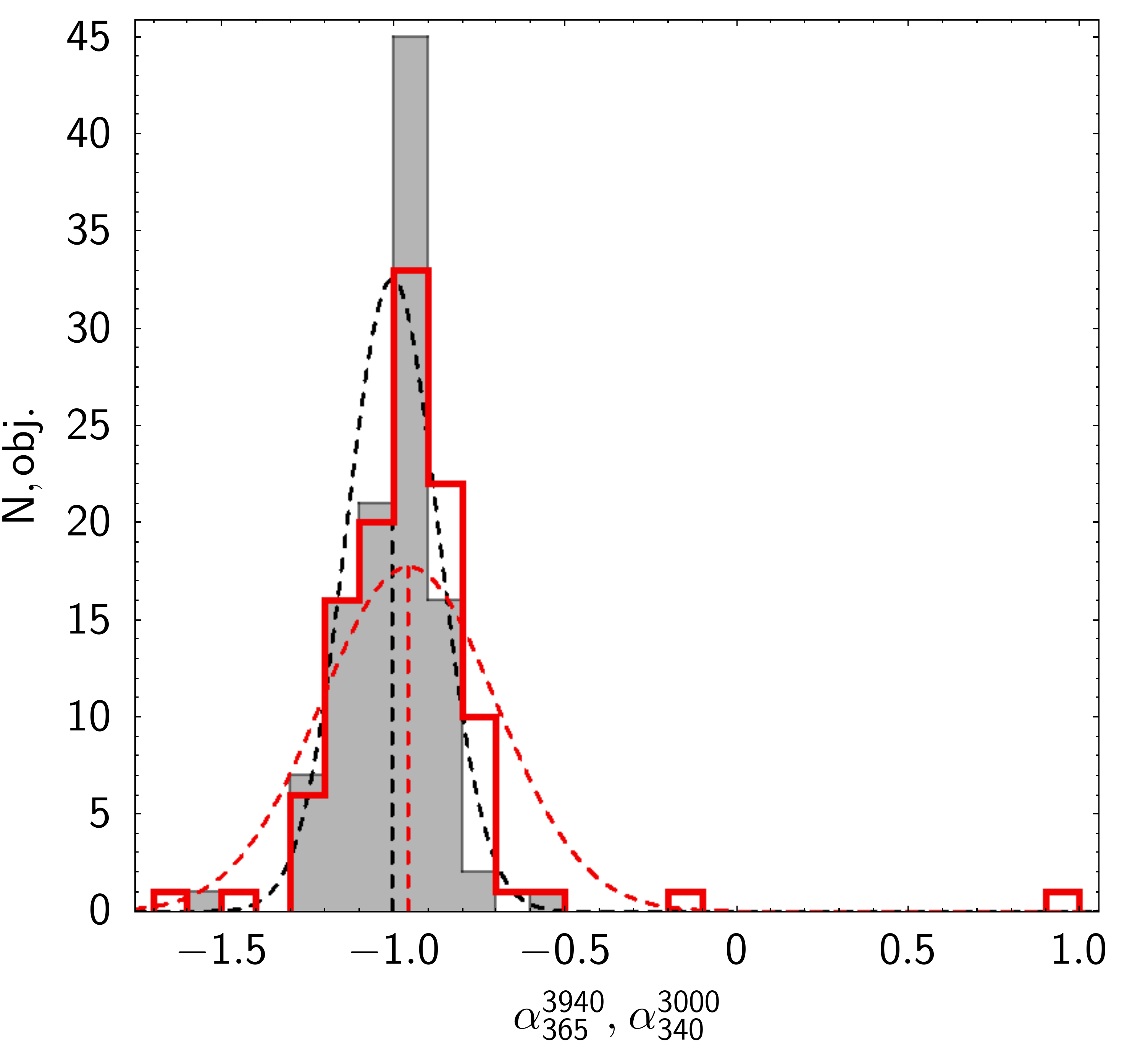}}
{b)\includegraphics[scale=0.23]{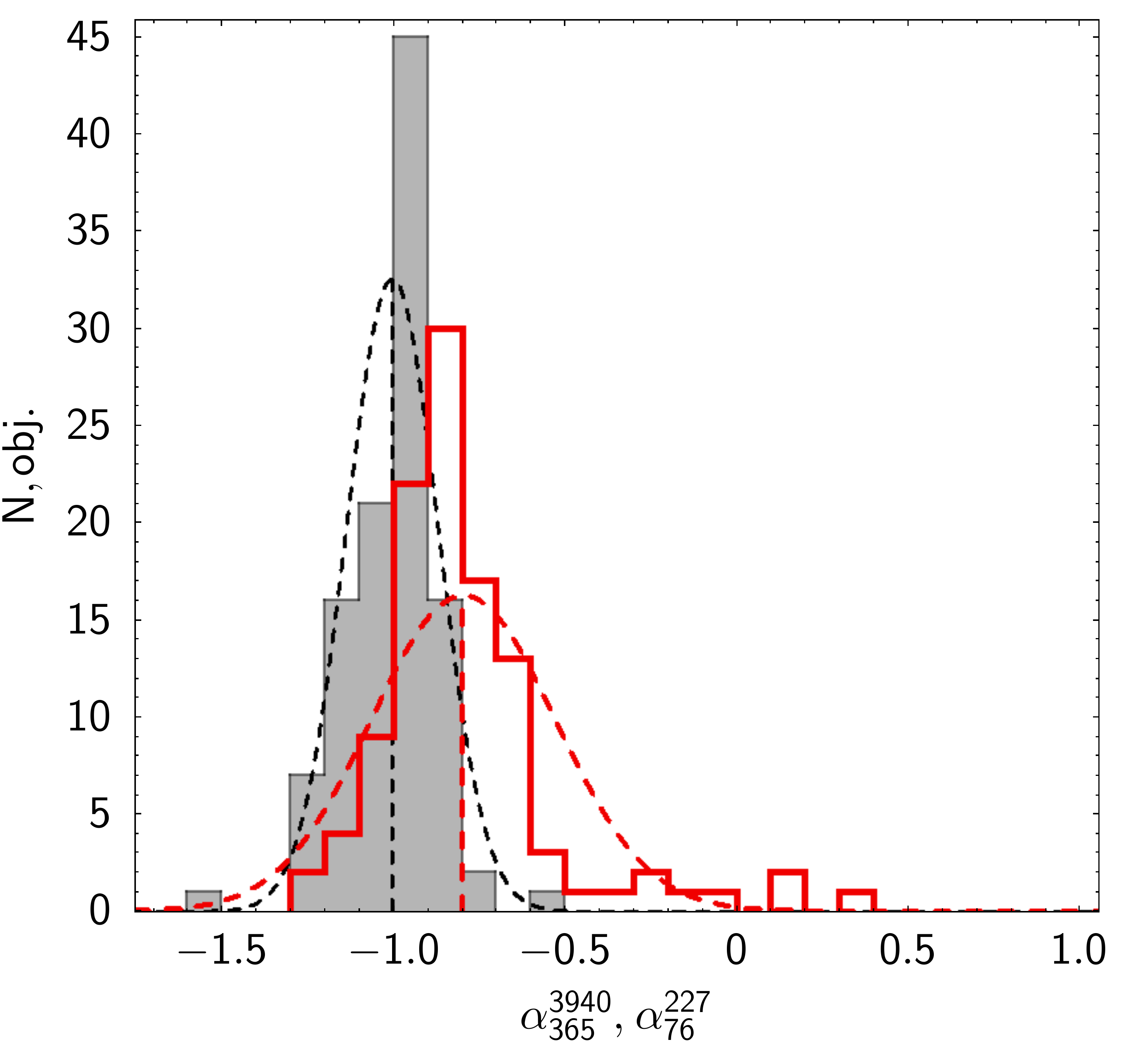}}
\caption{
Comparison of spectral indices $\alpha_{365}^{3940}$ calculated with linear approximation of ``old'' data, $\alpha_{340}^{3000}$ --  of ``new'' data, and $\alpha_{74}^{231}$ --  of ``glm'' data for radio sources of the Big Trio program. Distributions and their approximation by Gaussians for spectral indices: (a) $\alpha_{365}^{3940}$ (grey) and $\alpha_{340}^{3000}$ (red line); (b) distributions of spectral indices $\alpha_{365}^{3940}$ (grey) and $\alpha_{76}^{227}$ (red line).
}
\label{fig:Sp}
\end{figure}

The distributions of the ``old'' spectral indices $\alpha_{365}^{3940}$ (gray) and the ``new'' $\alpha_{340}^{3000}$ (red) are shown in the histograms of Fig.~\ref{fig:Sp}(a). The distributions of these indices do not differ significantly.

In contrast, when comparing the distributions of $\alpha_{365}^{3940}$ with $\alpha_{76}^{227}$ (see Fig.~\ref{fig:Sp}(b)) a noticeable shift is observed: the $\alpha_{76}^{227}$ indices tend to be flatter, indicating a systematic difference in the slope of the spectra at lower frequencies.

Note the decrease in the number of sources with steep and ultra-steep spectra when using newer data.
Thus, when calculating spectral indices using the ``old'' dataset from 112 sources of the sample, 90 (80\%) have $\alpha\leq -0.9$. Using the ``new'' dataset from 113 sources, 79 (70\%) meet this criterion.
However, when using the ``glm'' dataset, only 39 sources (35\%) have $\alpha\leq -0.9$.

\begin{figure}
\center
{a)\includegraphics[scale=0.23]{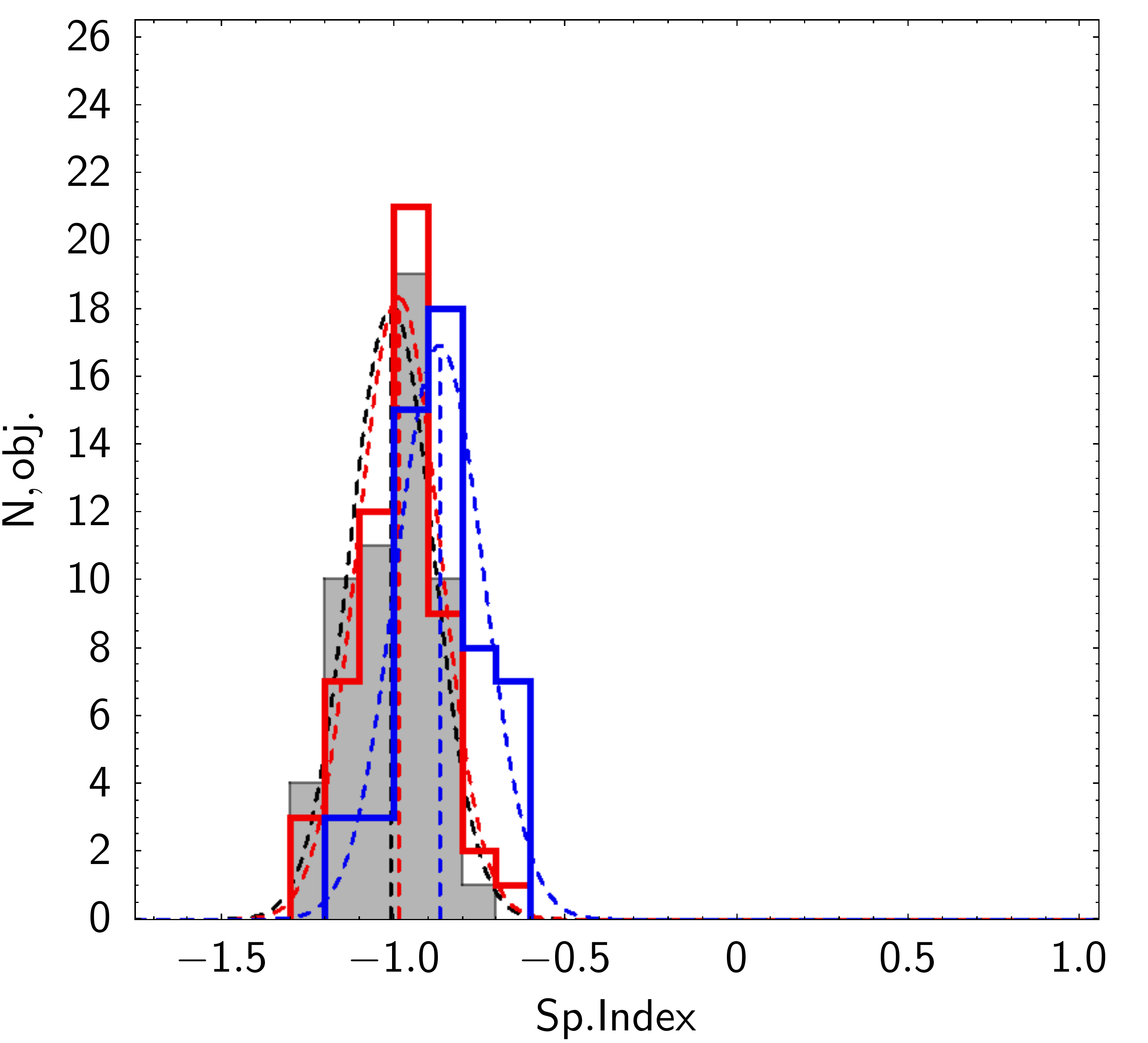}}
{b)\includegraphics[scale=0.23]{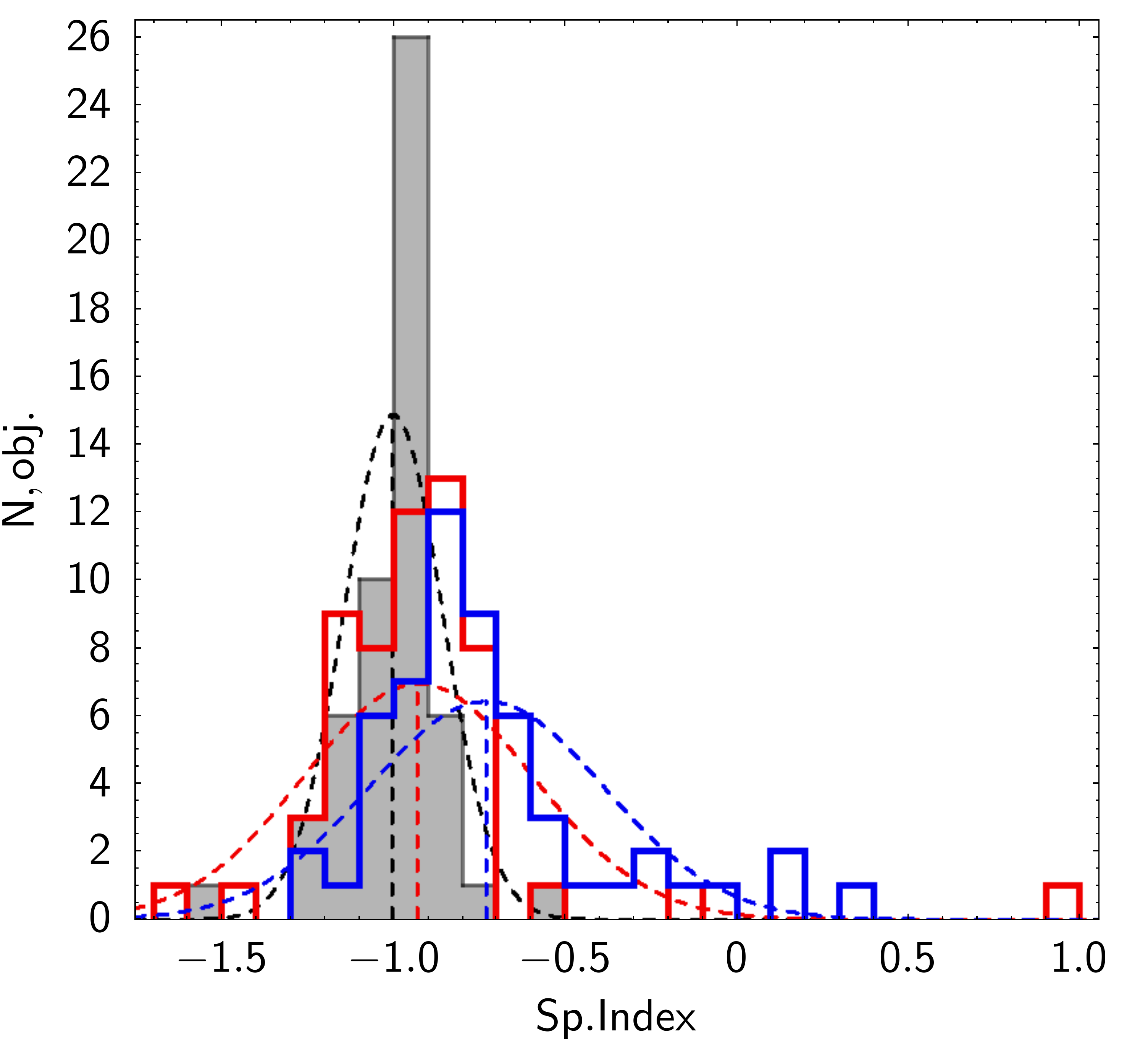}}
\caption{
Comparison of the distributions of spectral indices $\alpha_{365}^{3940}$ (gray), $\alpha_{340}^{3000}$ (red), and $\alpha_{76}^{227}$ (blue), along with their Gaussian approximations, is presented for two subsamples:  
(a) subsample I and  
(b) subsample II.
The histograms illustrate how the spectral slopes vary across datasets and frequency ranges, providing insight into the consistency and evolution of spectral properties within the sample.
}
\label{fig:Sp1}
\end{figure}

The relatively low percentage of steep-spectrum sources in the ``glm'' dataset can be attributed to spectral roll-off at low frequencies, likely caused by synchrotron aging of the electron population.

Next, we divided the radio sources into two subsamples:

\textbf{Subsample (I)} includes sources without a spectral peak, with spectral curvature parameter $SCP$ in the range $0.0\div0.5$~\citep{2011A&A...526A.148M}, and variability index $|V_{340}^{cc}| < 3$.  
For these sources, we assume that the linear approximation of the continuum spectrum is relatively insensitive to the choice of dataset used to determine the spectral index.  
This subsample contains 55 sources.

\textbf{Subsample (II)} comprises all remaining sources not included in Subsample I.
\begin{table}
\caption{Comparison of spectral indices of subsamples (I) and (II).\\ $Sp.Ind.$ -- mean and RMS of distribution; $Median$ -- median of distribution; $SS$ -- number of sources with steep spectra ($\alpha\leq-0.9$)}.
\centering
\begin{tabular}{|l|c|c|c|}
\hline
~~~~~&~$Sp.Ind.$~&~$Median$~&~$SS$ \\
\hline \hline
(I)  & $\alpha_{365}^{3940}=-1.05\pm0.10$ & -1.03 & 44 \\ 
(II) & $\alpha_{365}^{3940}=-1.04\pm0.12$ & -1.00 & 46 \\ 
\hline
(I)  & $\alpha_{340}^{3000}=-1.03\pm0.09$ & -1.01 & 43 \\ 
(II) & $\alpha_{340}^{3000}=-1.08\pm0.15$ & -1.06 & 36  \\ 
\hline
(I)  & $\alpha_{76}^{227}=-0.99\pm0.06$ & -0.96 & 22 \\
(II) & $\alpha_{76}^{227}=-1.04\pm0.09$ & -1.03 & 17 \\
\hline \hline
\end{tabular}
\label{tab:RCss}
\end{table}

The results of the comparison between the two subsamples are presented in Fig.~\ref{fig:Sp1} and Table~\ref{tab:RCss}.

When comparing subsamples (I) and (II), the discrepancy in the spectral index distributions is particularly pronounced for subsample (II) (see Fig.~\ref{fig:Sp1}b).  
In subsample (I), the number of sources with $\alpha \leq -0.9$, calculated using spectral indices from the ``old'' and ``new'' datasets, remained virtually unchanged (see Table~\ref{tab:RCss}).  
However, in subsample (II), the number of such sources decreased from 46 (based on the ``old'' data) to 36 (based on the ``new'' data).  
Additionally, the number of sources with $\alpha_{76}^{227} \leq -0.9$ is lower in subsample (II) than in subsample (I).

This can be explained by the nature of the sources in each group.  
Subsample (I) consists primarily of sources whose jets continue to be powered by the active nucleus, resulting in relatively stable spectral shapes.  
In contrast, subsample (II) includes variable sources, young radio sources with jets in early stages of development, and fading sources whose jets are no longer energized by the nucleus.  
As a result, subsample (II) demonstrates more pronounced changes in the continuum spectra compared to subsample (I), namely, a shift of the entire spectrum towards low frequencies and a decline at low frequencies.

\section{Results}
\textbf{\textit{Morphology.}}  
For the well-studied sample of steep-spectrum sources from the Big Trio program, we refined the radio morphology using high-resolution radio maps with angular resolutions ranging from $0.1^{\prime\prime}$ to $2.5^{\prime\prime}$.

Visual inspection of radio maps, along with cutouts from optical and infrared surveys, revealed that several objects previously considered single radio sources are, in fact, composed of two closely spaced sources on the sky. Six such cases were identified, four of which are confirmed as physically associated pairs based on host redshifts, with projected separations of approximately 60\,kpc.

The morphological classification of the sample is as follows:
FRII-type radio galaxies dominate the sample, comprising 72\%;
FRI-type galaxies account for 3\%;
Hybrid FRI/FRII sources make up 7\%;
Point sources unresolved in available radio maps constitute 18\%.

Among the double sources, 27\% exhibit a detectable radio core.  
Six sources show a core contribution exceeding 35\% and are classified as core-dominated triple (CDT) radio sources.  
Three sources display double-double (DD) lobes, indicative of restarted activity.  
Overall, 8\% of the sample shows morphological evidence of episodic activity in the radio domain.

Lobe deformation patterns suggest environmental interactions at varying distances from the active nucleus.
Winged lobes may result from jet reorientation due to accretion disk instability, possibly triggered by interaction with a nearby massive object.
Hybrid lobes are likely caused by environmental inhomogeneity.
Wide-angle tail (WAT) morphologies typically indicate the presence of the radio source within a galaxy group or cluster.

In total, we classified
12 sources as winged radio galaxies (WRGs),
8 sources as WATs,
4 sources as horseshoe-shaped or C-shaped.

Including an additional 8 hybrid FRI/FRII sources, approximately 20\% of the hosts in the sample\footnote{Some sources may exhibit mixed morphological features, e.g., DD and WAT.} have neighbors or are located in a group or cluster of galaxies.

The sample also includes two giant radio galaxies and 26 compact sources with linear sizes ranging from 0.7 to 19\,kpc.

\textbf{\textit{Variability.}}  
We compared total flux densities of radio sources at four frequency ranges: 340-365\,MHz, 1368-1400\,MHz, 3900-3940\,MHz, and 4850-5000\,MHz.

Among the 87 sources with available data in the VCSS and TXS catalogs, 44\% exhibit variability indices $|V_{340}| > 3$.  
In contrast, no variable sources were identified based on RACS and NVSS data.  
At higher frequencies, 31\% of 101 sources show $V_{3940} > 3$, and 37\% of 93 sources have $V_{4850} > 3$.

We examined the VCSS and TXS data in greater detail, as the time interval between their mean observation epochs spans approximately 40 years.  
The high proportion of sources with $|V_{340}| > 3$ is primarily attributed to differences in angular resolution between surveys, which affect the contribution of nearby sources to the total flux density $S_{365}$.  
Additionally, the underestimated values of $S_{340}$ for some double sources further contribute to elevated variability indices.

After applying corrections for these effects, 19 sources still exhibit $|V_{340}| > 3$.  
These may include sources with a spectral peak and compact, rapidly evolving objects whose spectra shift toward lower frequencies.  
We suggest that a time interval of approximately four decades is sufficient for such spectral evolution to become detectable.  
These sources represent about 11\% of the sample.

It is also possible that some sources possess extended, low surface brightness components that are undetected in the VCSS survey but are likely detectable in TXS and GLEAM.  
This hypothesis was not explored in detail in the present study.

Among the sources with $|V_{340}| > 3$, a subset may be genuinely variable.  
Approximately 10\% of the sample is suspected to be variable, primarily consisting of compact radio quasars.

\textbf{\textit{Continuum Spectra.}}  
We compared the spectral indices of the Big Trio sample sources derived from continuum spectra constructed by fitting flux density data obtained prior to 1996 and more recent data from modern radio surveys.

The lower proportion of steep-spectrum sources in the spectral index calculation based on GLEAM data is likely due to the addition  of low-frequency data, which allowed to refine the spectra in this region and identify spectral flattening or roll-off. However, we cannot rule out the possibility of internal evolution of some sources, leading to changes in the spectral shape, such as a shift of the spectral peak toward lower frequencies over time.

\acknowledgments{Observations were carried out using the RATAN-600 radio telescope. This research has made use of the NASA/IPAC Extragalactic Database (NED), operated by the Jet Propulsion Laboratory, California Institute of Technology, under contract with the National Aeronautics and Space Administration; and the CATS database, available via the Special Astrophysical Observatory website.}

\section*{Funding}  
This work was conducted within the framework of the state assignment of the Special Astrophysical Observatory of the Russian Academy of Sciences (SAO RAS), as approved by the Ministry of Science and Higher Education of the Russian Federation.

\bibliographystyle{aspb1}
\bibliography{Trio}

\end{document}